\documentclass[twocolumn,letterpaper,amsmath,amssymb,floatfix,aps,prb,superscriptaddress]{revtex4}

\usepackage{graphicx}
\usepackage{color}
\usepackage{dcolumn}  
\usepackage{bm}  
\usepackage{epsfig}

\begin{document}
\setlength\arraycolsep{2pt}

\date{\today}

\title{Many-body effects in van der Waals-Casimir interaction between graphene layers}

\author{Jalal Sarabadani}
\email{j.sarabadani@phys.ui.ac.ir}
\affiliation{Department of
Physics, University of Isfahan, Isfahan 81746, Iran}
\affiliation{Department of Theoretical Physics, J. Stefan
Institute, SI-1000 Ljubljana, Slovenia}

\author{Ali Naji}
\email{a.naji@damtp.cam.ac.uk}
\thanks{(corresponding author)}
\affiliation{Department of Applied Mathematics and Theoretical
Physics, Centre for Mathematical Sciences, University of
Cambridge, Cambridge CB3 0WA, United Kingdom}
\affiliation{School of Physics, Institute for Research in
Fundamental Sciences (IPM), Tehran 19395-5531, Iran}

\author{Reza Asgari}
\email{asgari@ipm.ir}
\affiliation{School of Physics, Institute for Research in
Fundamental Sciences (IPM), Tehran 19395-5531, Iran}

\author{Rudolf Podgornik}
\email{rudolf.podgornik@fmf.uni-lj.si}
\affiliation{Department of Theoretical Physics, J. Stefan
Institute, SI-1000 Ljubljana, Slovenia} 
\affiliation{Institute of Biophysics, School of Medicine and Department of Physics, Faculty
of Mathematics and Physics, University of Ljubljana, SI-1000
Ljubljana, Slovenia}

\begin{abstract}
Van der Waals-Casimir dispersion interactions between two apposed graphene layers, a graphene layer and a substrate, and 
in a multilamellar graphene system are analyzed within the framework of the Lifshitz theory. This formulation hinges on
a known form of the dielectric response function of an undoped or doped graphene sheet, assumed to be of a random phase
approximation form. In the geometry of two apposed layers the separation dependence of the van der Waals-Casimir interaction
for both types of graphene sheets  is determined and compared with some well known limiting cases. In a multilamellar array the
many-body effects are quantified and shown to increase the magnitude of the van der Waals-Casimir interactions.
\end{abstract}
\maketitle

\section{Introduction}
\label{section:intro}

Graphene appears to be the only known mono-atomic two-dimensional (2D) 
crystal and apart from the intrinsic interest it engenders, it is becoming more and more also
a focus of possible and desired advanced technological
applications~\cite{Novoselov}. It is for these reasons that in the
past several years we have witnessed a veritable explosion of
theoretical and experimental interest in graphene~\cite{Novoselov2}. The Nobel prize for physics in 2010 only
consolidated this trend. Graphene differs fundamentally from other
known 2D semiconductors because of its unique electronic band
structure,  viz. the monoatomic sheet of carbon atoms
arranged in a honeycomb lattice leads to an electron band
structure that displays quite unusual properties~\cite{Electronic}. The Fermi surface is reduced to just two points
in the Brillouin zone and the value of the band gap is reduced to
zero. The energy dispersion relation for both the conduction and
the valence bands are linear at low energy, namely less than 1 eV, meaning that the charge carriers behave as relativistic
particles with zero rest mass. The agent responsible for many of
the interesting electronic properties of graphene sheets is the
non-Bravais honeycomb-lattice arrangement of carbon atoms, which
leads to a gapless semiconductor with valence and conduction
$\pi$-bands.

States near the Fermi energy of a graphene sheet are
described by a massless Dirac equation which has chiral band
states in which the honeycomb-sublattice pseudospin is aligned
either parallel or opposite to the envelope function momentum.
The Dirac-like wave equation leads to both unusual
electron-electron interaction effects and to unusual response to
external potentials. When the graphene sheet is chemically doped
with either acceptor or donor impurities its carrier mobility can
be drastically decreased~\cite{Doped}. Because of its 2D periodic
structure graphene is closely related to single wall carbon
nanotubes, being in fact a carbon nanotube rolled out into a
single 2D sheet~\cite{Saito}. The main difference between the
electronic properties of single wall carbon nanotubes and graphene
is that the former show circumferential periodicity and curvature
that leave their imprint also in the electronic spectrum and
consequently also van der Waals (vdW) interactions~\cite{Rajter1,
Rajter2}.

On the other hand, graphite appears to be the poor cousin of
graphene though it is the stable form of carbon at ordinary
temperatures and pressures. Many efforts have been invested into
understanding its structural and electronic details (for an
account see Ref.~\cite{Graphitecalcs}). Various known modifications of
graphite differ primarily in the way the mono-atomic
two-dimensional graphene layers stack. Their stacking sequence in
terms of commonality is ABA for the Bernal structure, AAA for
simple hexagonal graphite or ABC for the rhombohedral graphite~\cite{Graphite}.

Graphene layers in graphitic systems are basically closed shell
systems and thus have no covalent bonding between layers which
makes them almost a perfect candidate to study long(er) ranged
non-bonding interactions. Indeed, they are stacked at an
equilibrium interlayer spacing of about 0.335 nm and are held
together primarily by the non-bonding long range vdW
interactions~\cite{GraphitevdW}.  Therefore the interaction
between graphene layers can be described as a balance between
attractive vdW dispersion forces and corrugated
repulsive (Pauli) overlap
forces \cite{Crespi}, 
following in this respect closely the paradigm of nano-scale
interactions~\cite{Rudi-RMP-2010}.

Besides a few notable exceptions~\cite{earlymanybodyeffects},
until 2009 many electronic and optical properties of graphene
could be explained within a single-particle picture in which
electron-electron interactions are completely neglected. The
discovery of the fractional quantum Hall effect in
graphene~\cite{fqhe} represents an important hallmark in this
context.  By now there is a large body of experimental
work~\cite{eeinteractionsgraphene, bostwick_science_2010, kotov_arXiv_2010}
 showing the relevance of electron-electron interactions in a
number of key properties of graphene samples of sufficiently high
quality.

Because of band chirality, the role of electron-electron
interactions in graphene sheets differs in some essential
ways~\cite{Asgari-MacDonald-PRL,ourdgastheory,dassarmadgastheory}
from the role which it plays in an ordinary 2D electron gas. One
important difference is that the contribution of exchange and
correlation to the chemical potential is an increasing rather than
a decreasing function of carrier density. This property implies
that exchange and correlation increases the effectiveness of
screening, in contrast to the usual case in which exchange and
correlation weakens screening~\cite{dft}. This unusual property
follows from the difference in sublattice pseudospin chirality
between the Dirac model's negative energy valence band states and
its conduction band
states~\cite{Asgari-MacDonald-PRL,ourdgastheory}, and in a uniform
graphene system is readily accounted for by many-body perturbation
theory.

\begin{figure}[b]
\begin{center}
\includegraphics[width=5cm]{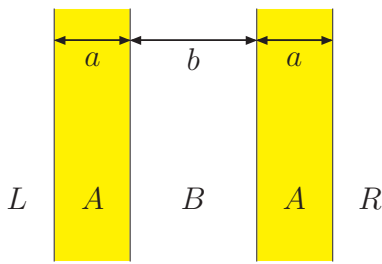}
\caption{ (Color online) Schematic presentation of two graphene
sheets of finite thickness immersed in vacuo at a separation of $b$.
The thickness of both (left and right) layers are equal to $a$.
 In view of later generalizations, we have labeled the left semi-infinite vacuum layer with $L$, the 
graphene layers with $A$, the intervening vacuum layer with $B$ and the
right vacuum layer with $R$. The form of the dielectric functions of the graphene layers is
given in Eq.~(\ref{eq:dielectric-function-graphene-sheet}).}
\label{fig:Schematic-two-graphene-sheet}
\end{center}
\end{figure}

In this work we focus our efforts on the vdW dispersion
component of the graphene stacking interaction. Dispersion forces
can be formulated on various levels~\cite{vdWgeneral} giving
mostly consistent results for their strength and separation
dependence. In the context of graphene stacking interactions,  the
problem can be decomposed into the calculation of the dielectric
response of the carbon sheets and the subsequent calculation of
the vdW interactions either via the quantum-field-theory-based Lifshitz approach, as advocated in this paper, by
means of the electron correlation energy~\cite{Sernelius} or the
non-local vdW density functional theory~\cite{Rydberg}.
One can show straightforwardly that in fact the non-local van der
Waals functional approach of the density functional theory and the
Lifshitz formalism are in general equivalent~\cite{Veble}.

Specifically we will calculate the vdW-Casimir interaction free
energy, per unit area between two graphene sheets as a function of the seperation between them,
in a system composed of
\begin{itemize}
\item[a)]  two apposed undoped or doped graphene sheets,
\item[b)]  an undoped or a doped graphene layer over a semi-infinite substrate, and
\item[c)]  a multilayer (infinite) array of graphene sheets.
\end{itemize}
In the latter case we will investigate the many-body non-pairwise additive effects
in the effective interaction between two sheets within a multilayer array. We should note
that non-pairwise additive effects are ubiquitous in the context of vdW
interactions~\cite{vdWgeneral} often leading to non-trivial properties of macromolecular
interactions. In this case they will lead to variations in the equilibrium stacking separation as a function of
the number of layers in a graphitic configuration. In the
calculation of the vdW-Casimir free energy we will
employ the dielectric response function of a single graphene layer
calculated previously~\cite{Asgari-MacDonald-PRL}.

\section{vdW-Casimir interaction between two layers of graphene}
\label{section:two-graphene-layers}

The geometry of the system composed of two parallel graphene
layers with thicknesses $a$, facing each other in a bilayer
arrangement at a separation $b$, is shown schematically in
Fig.~\ref{fig:Schematic-two-graphene-sheet}.
 In view of later generalizations we label the left semi-infinite
vacuum space as $L$, graphene sheets as $A$, the intervening layer
as $B$ and the right semi-infinite vacuum space by $R$.

In order to calculate the vdW-Casimir dispersion interaction free energy in the planar geometry we
use the approach of Ref.~\cite{Rudi-JCP-119-1070-2003} where it
has been calculated {\em exactly} for a multilayer planar geometry. The thus derived general
form of the interaction free energy per unit area includes
retardation effects and is therefore valid for any spacing between the
layers.

For the system which is shown schematically in
Fig.~\ref{fig:Schematic-two-graphene-sheet}, the vdW-Casimir interaction free energy is obtained in
the Lifshitz form as
%
\begin{equation}
\frac{F_{gg}(b)}{S} = k_B T \sum_{\bf Q}
{\sum_{n=0}^{\infty}}^\prime  \ln \bigg[ 1+
\frac{{\cal D}_1(\imath \xi_n)}{{\cal D}_2(\imath \xi_n)} e^{-2 b \kappa_{B}(\imath \xi_n)} \bigg],
\label{eq:Free-Energy-Two-sheets}
\end{equation}
where ${F}_{gg}(b)$ stands for the graphene-graphene
interaction free energy as a function of the layer spacing $b$ (normalized in such a way  that it tends to zero 
at infinite interlayer separation). In the above Lifshitz formula
the $\bf Q$ summation is over the transverse wave vector and the $n$ summation (where
the prime indicates that the $n=0$ term has a weight of $1/2$) is
over the imaginary Matsubara frequencies 
\begin{equation}
\xi_n=\frac{2 \pi n k_B T}{\hbar},
\end{equation} 
where $k_B$ is the Boltzman constant, $T$ is the
absolute temperature, and $\hbar$ is the Planck constant devided
by $2 \pi$. All the quantities in the bracket depend on $\bf Q$ as well as $\xi_n$.
%
%

The other quantities entering the Lifshitz formula are defined as
\begin{widetext}
\begin{eqnarray}
{\cal D}_1(\imath \xi_n) &=& \Delta_{BA}(\imath \xi_n) \Delta_{AB} (\imath \xi_n)+ e^{-2 a \kappa_{A}(\imath \xi_n)}
\Delta_{BA}(\imath \xi_n) \Delta_{RA}(\imath \xi_n) + \nonumber\\
&& \qquad \qquad \qquad \qquad +\, e^{-2 a \kappa_{A}(\imath \xi_n)} \Delta_{AL}(\imath \xi_n)
\Delta_{AB}(\imath \xi_n) + e^{-4 a \kappa_A} \Delta_{AL} \Delta_{RA}, \nonumber\\
{\cal D}_2(\imath \xi_n) &=& 1 + e^{-2 a \kappa_{A}(\imath \xi_n)} \Delta_{AB}(\imath \xi_n) \Delta_{RA}(\imath \xi_n) +
e^{-2 a \kappa_{A}(\imath \xi_n)} \Delta_{AL}(\imath \xi_n) \Delta_{AB}(\imath \xi_n) \nonumber\\
&& \qquad \qquad \qquad \qquad +\, e^{-4 a \kappa_A(\imath \xi_n)} \Delta_{AL}(\imath \xi_n) \Delta_{BA}(\imath \xi_n) \Delta_{AB}(\imath \xi_n)
\Delta_{RA}(\imath \xi_n) , \label{eq:alpha-Two-sheets}
\end{eqnarray}
\end{widetext}
with
\begin{equation}
\Delta_{i~i-1}(\imath \xi_n) = \frac{ \epsilon_i(\imath \xi_n) \kappa_{i-1}(\imath \xi_n) - \epsilon_{i-1}(\imath \xi_n)
\kappa_{i}(\imath \xi_n) }{\epsilon_i(\imath \xi_n) \kappa_{i-1}(\imath \xi_n) + \epsilon_{i-1}(\imath \xi_n) \kappa_{i}(\imath \xi_n)},
\label{eq:Delta-Two-sheets}
\end{equation}
%
%
%
where $\Delta_{i~i-1}$ quantifies the dielectric discontinuity between
homogeneous dielectric layers in the system, where a layer labeled
by $i-1$ is located to the left hand side of the layer labeled by
$i$ (for details see Ref.~\cite{Rudi-JCP-119-1070-2003}).
%
%
Also $\kappa_i(\imath \xi_n)$ for each electromagnetic field  mode within the material $i$  is given by
\begin{equation}
\kappa_{i}^{2} (\imath \xi_n) =  Q^2 + \frac{ \epsilon_i (\imath
\xi_n)  \mu_i (\imath \xi_n) \xi_n^2 }{c^2},
\label{eq:rho-Matsubara-Two-sheets}
\end{equation}
where $c$ is the speed of light in vacuo, $Q$ is the magnitude of
the transverse wave vector, and $\epsilon_i (\imath \xi_n)$ and
$\mu_i (\imath \xi_n)$ are the dielectric function and the magnetic
permeability of the $i$-th layer at imaginary frequencies, respectively.
For the sake of simplicity we assume that for all layers $\mu_i(\imath \xi_n)=1$ and that the dielectric function
for vacuum layers equals to $1$ for all frequencies.

Note that $\epsilon_i (\imath \xi)$ is standardly referred to as the vdW-London transform of the dielectric function
and is defined via the Kramers-Kronig relations as \cite{Wooten}
\begin{equation}
\epsilon(\imath\xi)=1+\frac{2}{\pi}\int_{0}^{\infty}\frac{\omega\epsilon''(\omega)}{\omega^{2}+\xi^{2}}d\omega.
\label{eq:KK_transform}
\end{equation}
It characterizes the magnitude of spontaneous electromagnetic fluctuations at frequency $\xi$.
In general $\epsilon(\imath\xi)$ is a real, monotonically decaying function of the imaginary
argument $\xi$ (for details see Parsegian's book in Ref.~\cite{vdWgeneral}).

In order to proceed one needs the vdW-London transform of the dielectric function of all layers in the system.
The detailed $Q$- and $\omega$-dependent form of the dielectric function for undoped and/or
doped graphene layers are introduced in Secs.~\ref{Two-undoped-graphene} and \ref{Two-doped-graphene}.

\subsection{Two undoped graphene layers}
\label{Two-undoped-graphene}
We employ the response function of a graphene layer from
Refs.~\cite{Asgari-MacDonald-PRL, others} which for doped graphene
assumes the form
\begin{eqnarray}
\chi (Q,\imath \xi_n,\mu \neq 0) &=& - \frac{g \mu}{2 \pi \hbar^2
v^2} -
\frac{g Q^2}{16 \hbar \sqrt{\xi_n^2 + v^2 Q^2}} + \nonumber\\
&& \hspace{-2.5cm} + \frac{g Q^2}{8 \pi \hbar \sqrt{\xi_n^2 + v^2
Q^2} }
\mathrm{Re} \bigg[ \arcsin\big( \frac{2 \mu + \imath \xi_n \hbar}{v Q \hbar} \big)\nonumber\\
&& \hspace{-0.5cm} + \frac{2 \mu + \imath \xi_n \hbar}{v Q \hbar}
\sqrt{1- \bigg(\frac{2 \mu + \imath \xi_n \hbar}{v Q} \bigg)^2 }
\bigg], \label{eq:Response-function-graphene-sheet}
\end{eqnarray}
where $g=4$, $v\approx10^6\, {\mathrm{m}}/{\mathrm{s}}$ is the
Fermi velocity in graphene layer and $\mu= \varepsilon_F = \hbar v
k_F$ is the chemical potential, $\varepsilon_F$ the Fermi energy,
and $\hbar k_F$ is the Fermi momentum, where $k_F= (4 \pi
\rho/g)^{1/2}$ and $\rho$ is the the average electron
density.

To begin with, we assume that two layers are decoupled and ignore the interlayer Coulomb interaction. The vdW-London dispersion transform of the dielectric function on the level of the {\em random phase
approximation} (RPA) is then given by
\begin{equation}
\epsilon (Q,\imath \xi_n) = 1 - V(Q) \chi (Q,\imath \xi_n,\mu \neq
0) , \label{eq:dielectric-function-graphene-sheet}
\end{equation}
where $V(Q)$ is the (transverse) 2D Fourier-Bessel transform of
the Coulomb potential, $V(Q)= \frac{2 \pi e^2 }{4 \pi \epsilon_0
\epsilon_m Q}$,
 $e$ is the electric charge of electron, $\epsilon_0$ is the
permittivity of the vacuum and $\epsilon_m$ is the average of
the dielectric constant for the surrounding media which is equal to 1
for vacuum.
In what follows we furthermore assume that to the lowest order the
dielectric properties of the graphene layers are not affected by the variation of the separation
between them.  This assumption is also consistent with
the Lifshitz theory that presumes complete independence of the dielectric response functions
of the interacting layers.

For undoped graphene layer, i.e. $\rho=0$, the
expression for the vdW-London dispersion transform of the dielectric function simplifies substantially and assumes the form
\begin{equation}
\epsilon (Q,\imath \xi_n) = 1 + \frac{\pi \alpha g c Q }{8
\epsilon_m v \sqrt{ (\frac{\xi_n}{v})^2 + Q^2 } } ,
\label{eq:epsilon-undoped-graphene-sheet}
\end{equation}
where $\alpha$ is the electromagnetic fine-structure constant
$\alpha= \frac{e^2}{4 \pi \epsilon_0 \hbar
c}\approx\frac{1}{137}$. It should be noted that, in general, a model going beyond the RPA is necessary in order to account for enhanced correlation effects that would be present in an undoped system~\cite{LA}. In this paper, however, we restrict ourselves to the
 RPA approximation and analyze its predictions in detail.
%
%
%
\begin{figure}
\begin{center}
\includegraphics[width=8.5cm]{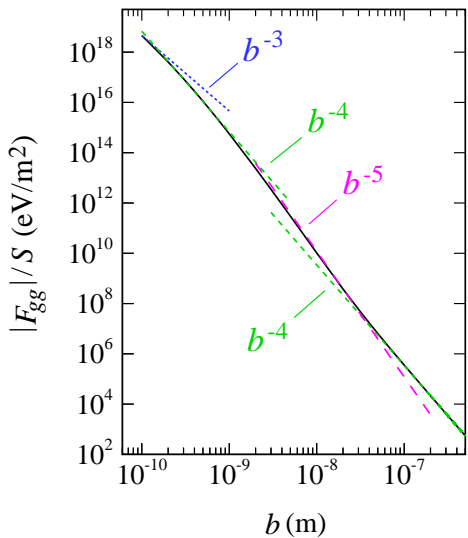}
\caption{ (Color online) Magnitude of the interaction free energy   per unit
area of the system composed of two undoped graphene layers
immersed in vacuo at an interlayer spacing $b$, and at temperature
$300~\mathrm{K}$. The functional dependence of the interaction
free energy on $b$ is compared with the following scaling forms: $b^{-3}, b^{-4}, b^{-5}$ and $b^{-4}$ as
the separation increases.}
\label{fig:Energy-two-undoped-graphene-sheet}
\end{center}
\end{figure}

The functional dependence of the interaction free energy of the
system per unit area, Eq.~(\ref{eq:Free-Energy-Two-sheets}), is
presented in Fig.~\ref{fig:Energy-two-undoped-graphene-sheet} as
a function of the separation between two graphene layers. We
assume that the graphene layers are immersed in vacuo and both of
them have the same thickness $1$\AA~as well as
equal susceptibilities. Note that in all cases considered in this paper 
 the interaction free energies as defined in Eq. (\ref{eq:Free-Energy-Two-sheets}) 
 are negative reflecting attractive vdW-Casimir force between graphene layers in vacuum. 
 For the sake presentation, we shall plot the absolute value (magnitude) of the free energy
 in all cases.

As one can discern from Fig.~\ref{fig:Energy-two-undoped-graphene-sheet} the
general dependence of the vdW-Casimir interaction free energy on the separation between the
graphene layers has the scaling form of a power law, $b^{-n}$, with a weakly varying 
separation-dependent scaling exponent, $n(b)$. This scaling exponent can be defined standardly as \cite{vdWgeneral}
\begin{equation}
n(b) = - \frac{d \ln{F_{gg}(b)}}{d \ln{b}}.
\label{powerdef}
\end{equation}
For two undoped graphene sheets we observe that  at small
separations the functional dependence of the free energy on
interlayer spacing yields the scaling exponent $n = 3$ for smallest values of the
separation. The scaling exponent  then steadily increases to $n = 4$, then $n = 5$,
and finally at asymptotically large separations it reverts back to $n = 4$. This
variation in the scaling exponent for the separation dependence of the
interaction free energy can be rationalized by invoking some well known results on the vdW  interaction in multilayer geometries (see e.g. the relevant discussions in  Ref.~\cite{vdWgeneral}).

For example, for two semi-infinite layers the interaction free
energy should go from the non-retarded form characterized by $n = 2$ for small
spacings, through retarded $n = 3$ form for larger spacings and then back to
zero-frequency-only form that also scales with $n = 2$ but with a
different prefactor than the non-retarded form. For two
infinitely thin sheets, on the other hand, we have the non-retarded
$n = 4$ form for small separation, followed by the retarded $n = 5$ form for
larger spacings and then reverting back to zero-frequency-only term with
$n = 4$ scaling, but again with a different prefactor than the
non-retarded limit. Furthermore, the transitions between various scaling forms
and the locations of the transition regions are not universal but depend crucially on
the characteristics of the dielectric spectra and can thus be quite complicated, sometimes
not yielding any easily discernible regimes with a quasi-constant scaling exponent $n$.

Reading Fig.~\ref{fig:Energy-two-undoped-graphene-sheet} with
this in mind we can come up with the following interpretation of
the calculated separation dependence: for small separation the
interaction free energy is dominated by the $n = 4$ dependence
except in the narrow interval very close to vanishing separation
where the dependence asymptotically levels off at $n < 4$ form.
This form is consistent with the non-retarded interaction between
two very thin layers, see above, for small but not vanishing separations. For
vanishing interlayer separations the final leveling-off of the scaling exponent  is due to the fact
that the system is approaching the limit of two semi-infinite layers where in principle $n \rightarrow 2$, but in
reality the finite thickness of the graphene sheets is way too small to observe this scaling
in its pure form. All we can claim is that for vanishing interlayer spacings the scaling
exponent drops below the $n = 4$ value, valid for two infinitely thin layers.

For larger values of the interlayer separations we then enter the retarded regime with $n = 5$
scaling exponent, again valid strictly for two infinitely thin layers. The
retarded regime finally gives way to the regime of asymptotically
large spacings where the interaction free energy limits towards
its form given by  the zero frequency term in the Matsubara summation
and characterized by $n = 4$ scaling dependence.  Obviously the
numerical coefficient in the small separation non-retarded and asymptotically large separation
regimes  (both with $n = 4$) are necessarily different.

The interaction free energy scaling with the interlayer separation
is thus completely consistent with the vdW-Casimir interactions between two thin
dielectric layers for all, except for vanishingly small, separations where finite thickness effects of the graphene sheets
leave their mark in a smaller value of the scaling exponent that should ideally approach the value valid for a
regime of interaction between two semi-infinite layers.

The numerical value of the interaction free energy per unit area at
$1~\mathrm{nm}$ is about $5.64\times 10^{14}~
\mathrm{eV}/\mathrm{m}^2$. That means that for two graphene layers
with surface area of $10^{-12}~\mathrm{m}^2$ the magnitude of
the free energy is about $564~\mathrm{eV}$ at 1 nm separation;
at $10~\mathrm{nm}$ it is about $0.01~\mathrm{eV}$ and at
$100~\mathrm{nm}$ about $3.6 \times 10^{-7}~\mathrm{eV}$ for
the same surface area.

\subsection{Two doped graphene layers}
\label{Two-doped-graphene}

For a doped graphene layer the vdW-London dispersion transform of the dielectric function can be read off
from Eqs.~(\ref{eq:dielectric-function-graphene-sheet}) and
(\ref{eq:Response-function-graphene-sheet}) as
\begin{eqnarray}
\epsilon (Q,\imath \xi_n,\mu \neq 0) &=& 1 + \frac{2 \pi \alpha c}{\epsilon_m Q v} \sqrt{\frac{\rho g}{\pi}} \nonumber\\
&& \hspace{-2.9cm} + \frac{\pi \alpha g c Q}{8 \epsilon_m v \sqrt{ (\frac{\xi_n}{v })^2 + Q^2}} \nonumber\\
&& \hspace{-2.9cm} - \frac{\alpha g c Q}{4 \epsilon_m v \sqrt{
(\frac{\xi_n}{v })^2 + Q^2} } \bigg \{ \arcsin \big[
\frac{1}{2} A_1 - \frac{1}{2} B_1 \big] \nonumber\\
&& \hspace{-2.9cm} + 2 \sqrt{\frac{4 \pi \rho/g}{Q}} \big( A_2^2+B_2^2 \big)^{1/4}
\cos \big[ \frac{1}{2} \arg (A_2 + \imath B_2) \big] \nonumber\\
&& \hspace{-2.9cm} - \frac{\xi_n }{v Q} \big( A_2^2+B_2^2
\big)^{1/4} \sin \big[ \frac{1}{2} \arg (A_2 + \imath B_2) \big]
\bigg\}, \label{eq:dielectric-function-Doped-graphene-sheet}
\end{eqnarray}
with the following coefficients
\begin{eqnarray}
A_1&=& \sqrt{\bigg(\frac{2 \sqrt{4 \pi \rho/g}}{Q}+1 \bigg)^2 + \bigg(\frac{\xi_n}{v Q} \bigg)^2}, \nonumber\\
B_1 &=& \sqrt{\bigg(\frac{2 \sqrt{4 \pi \rho/g}}{Q} -1 \bigg)^2 + \bigg(\frac{\xi_n}{v Q} \bigg)^2}, \nonumber\\
A_2&=&1+ \bigg( \frac{\xi_n}{v Q}\bigg)^2 - 16 \frac{\pi \rho }{g Q^2}, \nonumber\\
B_2&=& - 4 \frac{ \xi_n \sqrt{4 \pi \rho/g} }{v Q^2}.
\label{eq:dielectric-function-Doped-graphene-sheet-Parameters}
\end{eqnarray}
%

%
\begin{figure}
\begin{center}
\includegraphics[width=8.5cm]{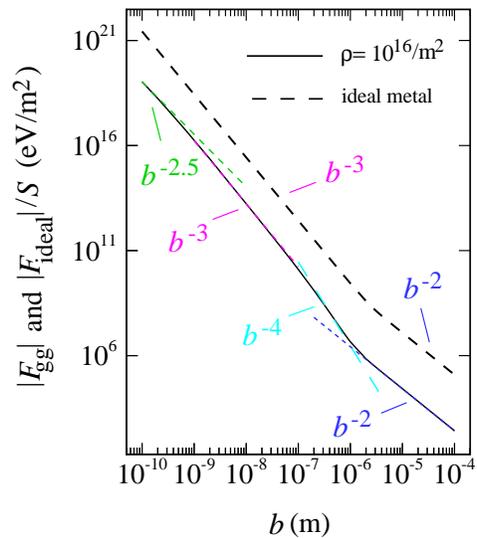}
\caption{ (Color online) Magnitude of the  interaction free energy   per unit
area of the system composed of two doped graphene layers (solid curve) compared with that of 
 two ideal metallic sheets (dashed curve)
immersed in vacuo as a function of the interlayer spacing $b$, at temperature
$300~\mathrm{K}$. The functional dependence of the free energy on
the interlayer spacing is compared with scaling forms $b^{-2.5}, b^{-3}, b^{-4}$
and $b^{-2}$ in various regimes of separation.}
\label{fig:Energy-two-Doped-graphene-sheet}
\end{center}
\end{figure}
With this dielectric response function we again evaluate the vdW-Casimir interaction free energy of
the system per unit area, Eq.~(\ref{eq:Free-Energy-Two-sheets}),
as a function of the separation between two graphene layers $b$ as
shown in Fig.~\ref{fig:Energy-two-Doped-graphene-sheet}. The interaction free energy
has again a scaling form with a scaling exponent varying with the separation between the layers.
It is clear that in this case it is much more difficult to partition the variation of the scaling exponent
into clear-cut piecewise constant regions.

As it can be seen from Fig.~\ref{fig:Energy-two-Doped-graphene-sheet} at small
separations the form of the functional dependence of the interaction
free energy has $n < 3$. Then for increasing spacings there follows a relatively broad regime with $n = 3-4$, followed
eventually by the scaling form with $n = 2$ for $b > 5 \times 10^{-6} \mathrm{m}$. One needs to add here that only the scaling
regime of $n = 2$ for asymptotically large separations and an intermediate regime
with $n = 3-4$ are clearly discernible.

We can gain some understanding of these regimes by comparing with the various exact limits in the
layer geometry as before. Such comparison is however not as straightforward as before. The asymptotic $n = 2$ regime is easiest to rationalize: it has the same scaling form as the {\em finite temperature} vdW-Casimir interaction between two metallic sheets at asymptotically large separations.
The presence of free charges would in fact be a reasonable characterization of doped graphene layers. For smaller separations we then enter the regime dominated by the retardation effects with $n \simeq 3-4$ form, and finally for vanishing separations we approach the regime of $n = \frac52$. Recent calculations of vdW interactions
between thin metallic layers indeed lead to exactly this exponent for small layer separations ~\cite{sernelius2}. 
The doped graphene sheet results would thus indicate that the dependence of the vdW-Casimir interactions free energy on the separation could be rationalized in terms of interactions between two thin metallic sheets.

For comparison we have also plotted the interaction free energy between two {\em ideal} metallic sheets 
which exhibits a much stronger attractive interaction free energy, i.e. \cite{vdWgeneral} 
\begin{eqnarray}
\frac{ {F}_{\mathrm{ideal}}}{S} &=& k_B T  {\sum_{n=- \infty}^{\infty}}
\int \frac{d^2  p}{(2 \pi)^2} \ln \big[ 1- e^{-2 b
\sqrt{p^2 + (\xi_n/c)^2}}\big] \nonumber \\
&=& - \frac{k_B T
\zeta (3)}{8 \pi b^2} +2 k_B T \times 
 \\ 
&&\hspace{1cm}\times 
\sum_{n=1}^{\infty} \int
\frac{d^2 p}{(2 \pi)^2} \ln \big[ 1- e^{-2 b \sqrt{p^2 +
(\xi_n/c)^2}}\big]. \nonumber
\end{eqnarray}

The numerical value of the energy per unit area at
$1~\mathrm{nm}$ is about $1.77\times 10^{16}~
\mathrm{eV}/\mathrm{m}^2$ which is equal to $1.77\times 10^{4}~
\mathrm{eV}$ for surface area of $10^{-12}~\mathrm{m}^2$; at
$10~\mathrm{nm}$ it is about $16.9~\mathrm{eV}$ and at
$100~\mathrm{nm}$ it is about $0.01~\mathrm{eV}$ for the same
surface area.

%
\begin{figure}
\begin{center}
\includegraphics[width=8.5cm]{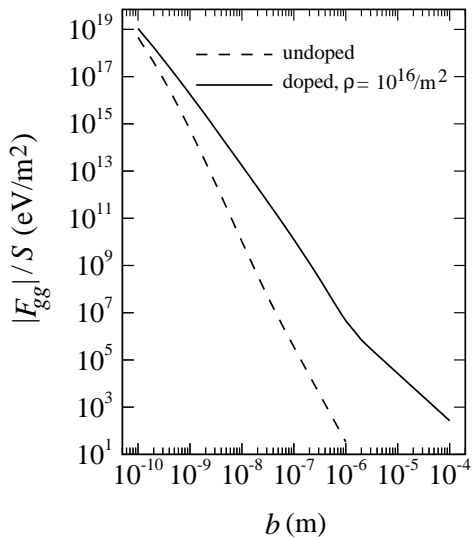}
\caption{Magnitude of the free energy   per unit
area of the system composed of two undoped (dashed line) and doped
(solid line) graphene layers (with the electron density $\rho=
10^{16} /\mathrm{m}^2$) immersed in vacuo as a function of the
interlayer spacing $b$, at temperature $300~\mathrm{K}$. As seen for all separations the magnitude of the interaction free
energy for doped graphene layers is greater than that of the
undoped one.
} \label{fig:Energy-two-undoped-Doped-graphene-Comparison}
\end{center}
\end{figure}

\begin{figure}
\begin{center}
\includegraphics[width=8.5cm]{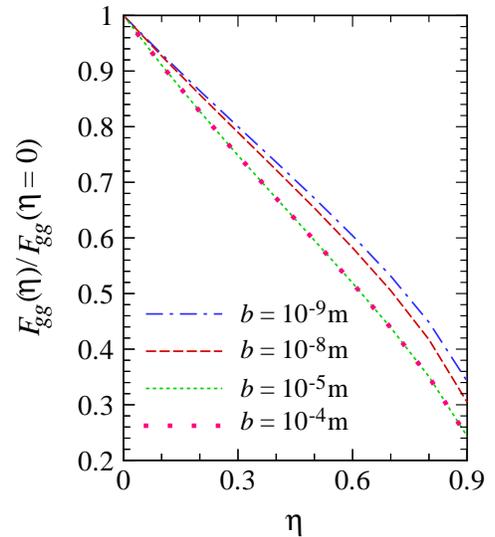}
\caption{(Color online) The rescaled interaction free energy, $F_{gg} (\eta) / F_{gg}
(\eta=0)$, for the system composed of two doped graphene layers
with different electron densities immersed in vacuum as a
function of $\eta= \frac{\rho_1
-\rho_2}{\rho_1+\rho_2}$ for different interlayer
separations  $b=1~\mathrm{nm}, 10~\mathrm{nm}, 10
~\mu\mathrm{m},$ and $100~\mu\mathrm{m}$ (from top to bottom)
at temperature $300~\mathrm{K}$. Note that $F_{gg} (\eta=0)$ has been calculated for $\rho_1 = \rho_2 = 10^{16} / {\mathrm{m}}^2 $. }
\label{fig:Energy-two-Different-Doped-graphene}
\end{center}
\end{figure}

\subsection{Doped vs. undoped graphene}

It is instructive to compare the interaction between two graphene
layers in the undoped and doped cases. For this purpose we have
plotted the interaction free energy of the system for both cases
in Fig.~\ref{fig:Energy-two-undoped-Doped-graphene-Comparison}.
The electron density in doped graphene is assumed to be
$\rho=10^{16}/\mathrm{m}^2$ (solid curve). The dashed curve is
the interaction free energy of the system composed of two undoped
graphene layers. As it can be seen, for all separations the magnitude
of the interaction free energy for doped graphene layers is more
than that of the undoped one (note again that the interaction 
free energies as defined in Eq. (\ref{eq:Free-Energy-Two-sheets}) are negative in the present case due to attractive 
vdW-Casimir force between graphene layers in vacuum). At separation $1~\mathrm{nm}$ the vdW-Casimir interactions
for doped graphene is about 30 times the magnitude of the
interaction for undoped graphene, while at the separation of $10~
\mathrm{nm}$ this ratio is about 1600. This means that the
attractive interaction between graphene layers is enhanced when the
contribution of the electron density in the dielectric function of
the graphene layers is taken into account. This same trend was observed also in the
work of Sernelius~\cite{Sernelius} and is clearly a consequence of the fact that
the largest value of vdW-Casimir interactions is obtained for ideally polarizable, i.e. metallic
layers. The closer the system is to this idealized case, the larger the corresponding vdW-Casimir
interaction will be.

Let us investigate also the effect of asymmetry of doped graphene sheets on vdW-Casimir interactions between them.
Introducing the dimensionless parameter $\eta= \frac{\rho_1 -\rho_2}{\rho_1 + \rho_2}$,
where $\rho_i$ is the electron density of the $i$-th graphene layer ($i=1,2$), we find
that the interaction free energy depends on the asymmetry in the system. When
$\eta = 0$, the electron densities are the same for both graphene
layers and we have a symmetric case, whereas $\eta=1$ means that one of the graphene layers is
undoped while the other one is doped, leading to an asymmetric case. In
Fig.~\ref{fig:Energy-two-Different-Doped-graphene} we have
plotted the rescaled free energy $F_{gg} (\eta) / F_{gg}(\eta=0)$ of the system composed of two graphene layers as a
function of $\eta$ for different values of the interlayer separation. The magnitude of the electron density for one of the layers
has been fixed at $\rho_1 = 10^{16} /{\mathrm{m}}^2 $ while that of the other layer, $\rho_2$, varies. As
seen from this figure, the curves show a monotonic dependence on $\eta$  with a
stronger interaction at smaller values of $\eta$. Note that at large
separations the curves tend to coincide  and will become indistinguishable.
The asymmetry effects are therefore largest at small separations between the interacting graphene layers.

\begin{figure}[t]
    \begin{center}
        \includegraphics[width=8.5cm]{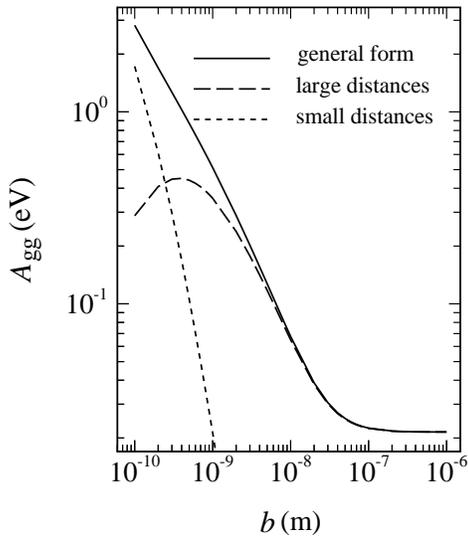}
\caption{Hamaker coefficient for a system
of two undoped  graphene layers as a function
of the distance between them, $b$. The top solid line shows the
general form of the Hamaker coefficient as defined via
Eq.~(\ref{eq:Hamaker-coefficient-general-form-two-graphene-layers}),
the dashed line shows the limiting large-distance form, Eq.~(\ref{eq:Hamaker-coefficient-large-distances-two-graphene-layers}),
and the dotted line shows the limiting small-distance form, Eq.~(\ref{eq:Hamaker-coefficient-small-distances-two-graphene-layers}). }
\label{Hamaker-two-graphene-layers}
\end{center}
\end{figure}

\subsection{Hamaker coefficient for two graphene layers}

The general form of the Hamaker coefficient, $A_{gg}$,  for a system composed of
two dielectric layers of finite thickness (Fig.~\ref{fig:Schematic-two-graphene-sheet}) when retardation effects
are neglected is defined via \cite{vdWgeneral}
\begin{equation}
\frac{F_{gg} (b)}{S} = - \frac{A_{gg}}{12 \pi b^2} \left[ 1
- \frac{2 b^2}{(b+a)^2} + \frac{b^2}{(b+ 2a)^2} \right]. 
\label{eq:Hamaker-coefficient-general-form-two-graphene-layers}
\end{equation}
At large separations $b \gg a$, the Hamaker coefficient, $A^{\mathrm{large}}_{gg}$,
can be obtained from
\begin{equation}
\frac{F_{gg} (b)}{S} = - \frac{A^{\mathrm{large}}_{gg}
a^2}{2 \pi b^4}  ,
\label{eq:Hamaker-coefficient-large-distances-two-graphene-layers}
\end{equation}
while at small separations $b \ll a$, it can be
read off from
\begin{equation}
\frac{F_{gg} (b)}{S} = - \frac{A^{\mathrm{small}}_{gg} }{12
\pi b^2} .
\label{eq:Hamaker-coefficient-small-distances-two-graphene-layers}
\end{equation}
In Fig.~\ref{Hamaker-two-graphene-layers}, we show
the Hamaker coefficient as a function of the layer separation for a system of two undoped  graphene layers. 
We show the general form of the Hamaker coefficient
from Eq. (\ref{eq:Hamaker-coefficient-general-form-two-graphene-layers}) (solid line) as well as
the limiting forms at small (dotted line, Eq. (\ref{eq:Hamaker-coefficient-small-distances-two-graphene-layers})) and large
(dashed line, Eq. (\ref{eq:Hamaker-coefficient-large-distances-two-graphene-layers})) separations. The large-distance limiting form
obviously coincides with the general form at separations beyond $50~\mathrm{nm}$. The small-distance limiting form tends to the
general form at small separations but given that   the thickness of the layers is only about $1$\AA, it is expected to merge with
the general form in sub-{\AA}ngstr\"om separations. For doped graphene, a similar analysis to Eq. (\ref{eq:Hamaker-coefficient-general-form-two-graphene-layers}) is not possible because the corresponding general expression for the interaction free energy is missing. 
%
%

\section{vdW-Casimir interaction between a graphene layer and a semi-infinite substrate}

\begin{figure}
\begin{center}
\includegraphics[width=5.0cm]{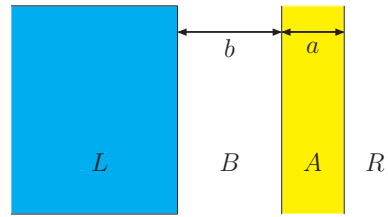}
\caption{ (Color online) Schematic presentation of a  graphene layer of thickness $a$ (labeled by $A$) apposed to
a substrate (labeled by $L$) at a separation $b$.  We have labeled the
intervening layer (assumed to be vacuum) with $B$, and the right
one with $R$. The dielectric function of the graphene layer is
defined via Eq.~(\ref{eq:dielectric-function-graphene-sheet})
and for substrate via Eq.~(\ref{eq:epsilon-substrate}).}
\label{fig:Schematic-substrate-graphene-sheet}
\end{center}
\end{figure}

In this section we study the interaction between a graphene layer
and a semi-infinite dielectric substrate as depicted schematically in
Fig.~\ref{fig:Schematic-substrate-graphene-sheet}. For this
system the free energy per unit area is
%
\begin{eqnarray}
\frac{F_{sg}(b)}{S} = && k_B T \times \nonumber\\
&& \hspace{-0.7cm} \times \sum_{{\bf Q}} {\sum_{n=0}^{\infty}}^{\prime}
\ln \bigg[ 1+ \frac{ \Delta_{AB} + \Delta_{RA} e^{-2 a \kappa_A}
}{1+\Delta_{AB} \Delta_{RA} e^{-2 a \kappa_A}} \Delta_{BL} e^{-2 b
\kappa_{B}} \bigg],\nonumber\\
\label{eq:Free-Energy-substrate-and-Graphen-sheet}
\end{eqnarray}
where ${F}_{sg}$ now stands for the interaction free energy
between the substrate and the graphene layer. We have excluded the explicit dependence of
the quantities in the bracket on the imaginary Matsubara frequencies, but they are the same as in Eq.~(\ref{eq:Delta-Two-sheets}).

%

In order to gain insight into the magnitude of the vdW-Casimir interaction free energy and for the sake of simplicity, we assume that the semi-infinite substrate is made of SiO$_2$ which has the
vdW-London dispersion transform of the dielectric function of the form \cite{Ninham-Mahanty-Book,vdWgeneral}
\begin{equation}
\epsilon_L (\imath \xi_n)
=1+\frac{\mathrm{C}_{\mathrm{UV}}\omega_{\mathrm{UV}}^2}{\xi_n^2
+\omega_{\mathrm{UV}}^2}+\frac{\mathrm{C}_{\mathrm{IR}}\omega_{\mathrm{IR}}^2}{\xi_n^2
+\omega_{\mathrm{IR}}^2}, \label{eq:epsilon-substrate}
\end{equation}
where the values of the parameters, C$_{\mathrm{UV}}$ = 1.098, C$_{\mathrm{IR}}$ = 1.703,
$\omega_{\mathrm{UV}} = 2.033 \times 10^{16}$ rad/s, and
$\omega_{\mathrm{IR}} = 1.88 \times 10^{14}$ rad/s have been determined from a
fit to optical data \cite{Hough}. The static dielectric permittivity of SiO$_2$ is then obtained as $ \epsilon(0) = 3.81$.
A characteristic feature of the vdW-London transform of SiO$_2$ is thus that it
contains two relaxation mechanisms. The first one is due to electronic polarization and the second one is due to
ionic polarization.  All calculations of the vdW-Casimir interaction free energy
are done at room temperature ($300~\mathrm{K}$).

%
\begin{figure}
\begin{center}
\includegraphics[width=8.5cm]{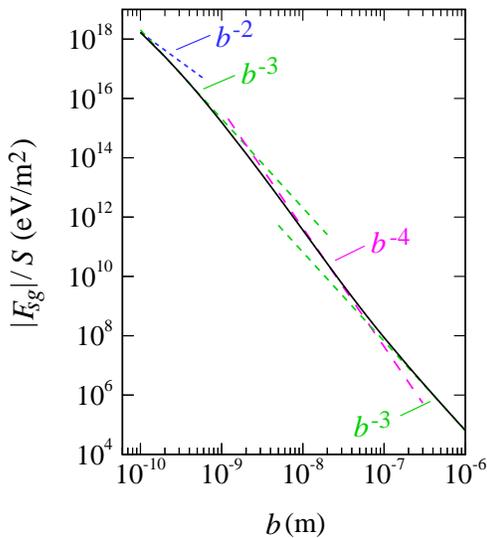}
\caption{ (Color online) Magnitude of the interaction free energy of the
system composed of a SiO$_2$ substrate and an undoped graphene
layer, Eq.~(\ref{eq:Free-Energy-substrate-and-Graphen-sheet}),
plotted at temperature $300~\mathrm{K}$ as a function of the
separation $b$. The functional dependence of the free energy on
 $b$ is compared with scaling forms $b^{-2}, b^{-3}, b^{-4}$
and $b^{-3}$ in various regimes of separation.}
\label{fig:Energy-Substrate-undoped-Graphene-Slope}
\end{center}
\end{figure}
%

%
\begin{figure}[t]
\begin{center}
\includegraphics[width=8.5cm]{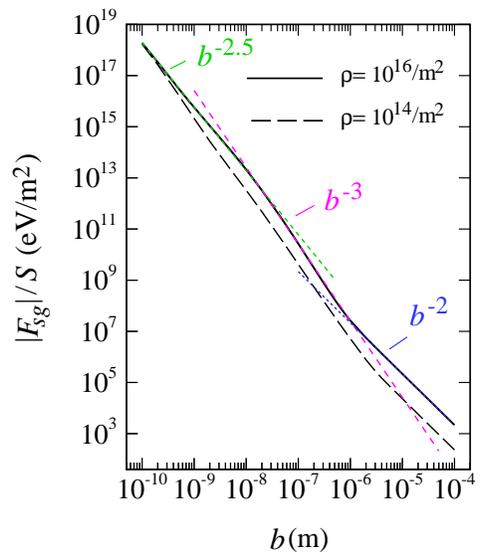}
\caption{ (Color online) Magnitude of the interaction free energy of the
system composed of a SiO$_2$ substrate and a doped graphene layer,
Eq.~(\ref{eq:Free-Energy-substrate-and-Graphen-sheet}) is
plotted at temperature $300~\mathrm{K}$ as a function of the
separation $b$. Here the functional dependence of the interaction
free energy for the graphene doping electron density $\rho = 10^{16} /\mathrm{m}^2$
(solid line)  is compared with scaling forms $b^{-2.5}, b^{-3}$
and $b^{-2}$ in various regimes of separation.
We also include the same plot for the doping electron density
$\rho = 10^{14} /\mathrm{m}^2$ (dashed line). }
\label{fig:Energy-Substrate-doped-Graphene-Slope}
\end{center}
\end{figure}

%
\subsection{Undoped graphene apposed to a substrate}

Using the vdW-London transforms of the  dielectric functions  given in the preceding sections,
one can now calculate the free energy, Eq.~(\ref{eq:Free-Energy-substrate-and-Graphen-sheet}),  for an undoped graphene layer
next to a semi-infinite SiO$_2$ substrate. The results are shown in Fig.~\ref{fig:Energy-Substrate-undoped-Graphene-Slope}.


At small separations, the free energy varies with a scaling exponent
$n = 2$, while at larger separations one can distinguish the scaling regimes  $n = 3$, and
$n = 4$, finally approaching the asymptotic limit at large separations with $n = 3$. This sequence of interaction free
energy scalings can be rationalized as follows: at asymptotically large
separations we are at the zero-frequency Matsubara term for a semi-infinite layer and a thin sheet.
This case is right in between the asymptotically large separation limit for two semi-infinite layers ($n = 2$) and two
infinitely thin layers ($n = 4$). For smaller spacings we then progressively detect
contributions from higher Matsubara terms which lead to scaling exponent $n = 4$ that corresponds to
a retarded form of the interaction free energy and then for yet
smaller spacings the scaling exponent reverts to $n = 3$ non-retarded form of the vdW interaction
between a semi-infinite substrate and an infinitely thin sheet. For vanishing spacings the finite thickness of the sheet starts
playing a role and eventually,  we approach the $n = 2$ scaling for two semi-infinite layers.

The magnitude of the interaction free energy per unit area at $1~\mathrm{nm}$ is
about $1.57\times 10^{15}~\mathrm{eV}/\mathrm{m}^2$. It means
that for a system with surface area of $10^{-12}~\mathrm{m}^2$
this value is about $1.57 \times 10^{3}~\mathrm{eV}$. At
$10~\mathrm{nm}$ it is about $0.36~\mathrm{eV}$ and at
$100~\mathrm{nm}$ it is about $8.71\times 10^{-5}~\mathrm{eV}$
for the same surface area.

\begin{figure}[t]
    \begin{center}
        \includegraphics[width=8.5cm]{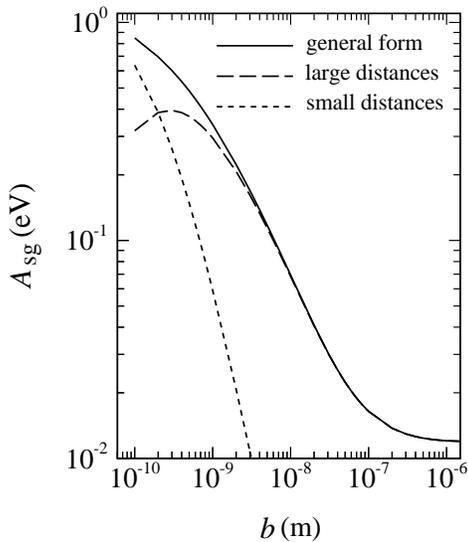}
\caption{Hamaker coefficient for a system of an undoped graphene layer apposed to a SiO$_2$ substrate
 as a function of the distance between them, $b$. The top solid line shows the
general form of the Hamaker coefficient as defined via
Eq.~(\ref{eq:Hamaker-coefficient-general-form-substrate-graphene}),
the dashed line show the limiting large-distance form, Eq.~(\ref{eq:Hamaker-coefficient-large-distances-substrate-graphene}),
and the dotted line shows the limiting small-distance form, Eq.~(\ref{eq:Hamaker-coefficient-small-distances-substrate-graphene}).}
\label{Hamaker-substrate-and-graphene}
\end{center}
\end{figure}

\subsection{Doped graphene apposed to a substrate}
The vdW-Casimir interaction free energy Eq.~(\ref{eq:Free-Energy-substrate-and-Graphen-sheet}) for a system composed
of a doped graphene layer next to a semi-infinite SiO$_2$ substrate is shown  Fig.~\ref{fig:Energy-Substrate-doped-Graphene-Slope} for two values of the electron density $\rho = 10^{14}/\mathrm{m}^2$
(dashed line) and $\rho = 10^{16}/\mathrm{m}^2$ (solid line).

At small separations the free energy shows the $n = \frac52$ scaling, while
for larger spacings it shows a scaling exponent $n = 3$, approaching the $n \rightarrow 2$ limit for asymptotically large separations.  In this respect the case of a thin doped graphene sheet apposed to a semi-infinite substrate
is very similar to the case of two thin doped layers, except that the retarded regime covers a smaller interval of spacings.
This means that the metallic nature of one of the interacting surfaces is enough to switch the behavior of the interaction
free energy completely towards the case of two metallic interacting surfaces. This case has in fact not yet been
thoroughly discussed in the literature. The changes in the slope appear to occur at the same values of the interlayer
spacing when the electron density decreases  (compare dashed and solid curves).


The magnitude of the free energy for the electron density $\rho
= 10^{14}/\mathrm{m}^2$ (dashed line) and surface area of
$10^{-12}~\mathrm{m}^2$ at $1~\mathrm{nm}$ is about
$2.12\times 10^{3}~\mathrm{eV}$, while at $10~\mathrm{nm}$ it
is about $3.28~\mathrm{eV}$, and at $100~\mathrm{nm}$ is
about $4.20\times 10^{-3}~\mathrm{eV}$.

These values increase for larger electron densities, e.g., for
$\rho = 10^{16}/\mathrm{m}^2$ (solid line) and the same surface
area, the free energy magnitude is $5.77\times 10^{3}~
\mathrm{eV}$ at $1~\mathrm{nm}$, while it is about $21.1~
\mathrm{eV}$ at $10~\mathrm{nm}$  and  about $2.67\times
10^{-2}~\mathrm{eV}$ at $100~\mathrm{nm}$.


\subsection{Hamaker coefficient for the graphene-substrate system}

The general form of the Hamaker coefficient, $A_{sg}$, for a system composed of
a dielectric layer of finite thickness apposed to a semi-infinite dielectric substrate (Fig.~\ref{fig:Schematic-substrate-graphene-sheet}) when retardation effects are neglected is defined via \cite{vdWgeneral}
\begin{equation}
\frac{F_{sg} (b)}{S} = - \frac{A_{sg}}{12 \pi b^2} \left[1 -
\frac{b^2}{(b+a)^2} \right]. 
\label{eq:Hamaker-coefficient-general-form-substrate-graphene}
\end{equation}
At large separations $b \gg a$, the Hamaker coefficient, $A^{\mathrm{large}}_{sg}$,
can be obtained from
\begin{equation}
\frac{F_{sg} (b)}{S} = - \frac{A^{\mathrm{large}}_{sg} a}{6
\pi b^3}  ,
\label{eq:Hamaker-coefficient-large-distances-substrate-graphene}
\end{equation}
while at small separations $b \ll a$, it can be
read off from
\begin{equation}
\frac{F_{sg} (b)}{S} = - \frac{A^{\mathrm{small}}_{sg} }{12
\pi b^2} .
\label{eq:Hamaker-coefficient-small-distances-substrate-graphene}
\end{equation}
In Fig.~\ref{Hamaker-substrate-and-graphene}, we again show
the Hamaker coefficient as a function of the layer separation for a system comprising a SiO$_2$ substrate and an undoped graphene layer. We show the general form of the Hamaker coefficient
from Eq. (\ref{eq:Hamaker-coefficient-general-form-substrate-graphene}) (solid line) as well as
the limiting forms at large (dashed line, Eq. (\ref{eq:Hamaker-coefficient-large-distances-substrate-graphene})) and small
(dotted line, Eq. (\ref{eq:Hamaker-coefficient-small-distances-substrate-graphene})) separations. The large-distance limiting form
obviously coincides with the general form at separations beyond $10~\mathrm{nm}$ but again since the thickness of the
graphene layer is only about $1$\AA, the  small-distance limiting form  is expected to merge with
the general form in sub-{\AA}ngstr\"om separations.
%
%


\section{vdW-Casimir interaction in a system composed of $N+1$ layers of
graphene} \label{Multilayers}

\begin{figure}[t] 
\vspace{+0.5cm}
\begin{center}
\includegraphics[width=8.5cm]{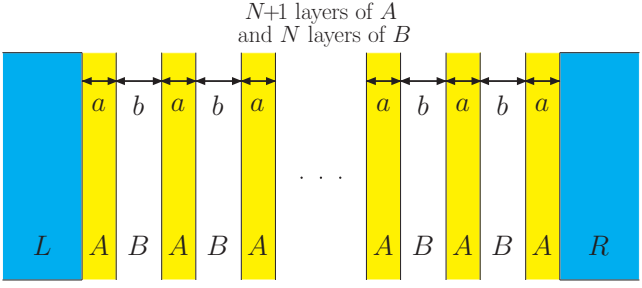}
\caption{(Color online) Schematic picture of the system
composed of $N+1$ layers of graphene ($A$) with equal thicknesses  $a$
 separated from each other by $N$ layers of vacuum ($B$)
with equal thicknesses of $b$. The two semi-infinite substrates in the left and right end of the system
are labeled by $L$ and $R$ and will be assumed to be vacuum. } \label{fig:N-layer-New}
\end{center}
\end{figure}

In this section we shall use the Lifshitz formalism in order to
study the many-body vdW interactions in a system composed of $N+1$
layers of graphene. The layers are separated from each other by
$N$ layers of vacuum and are bounded at the two ends by two
semi-infinite dielectric slabs as depicted in
Fig.~\ref{fig:N-layer-New}. The thickness of each graphene layer
is $a$ while the separation between two successive layers is $b$.
We have labeled the left semi-infinite dielectric medium with $L$,
the right one with $R$ (which both will be assumed to be vacuum), the graphene layers with $A$ and the
vacuum layers with $B$. Following
Ref.~\cite{Rudi-JCP-124-044709-2006},  one can calculate the
vdW-Casimir part of the interaction free energy, $F_N(a,b)$, in an explicit
form for any finite $N$. Interestingly, it turns out that for very
large values of $N$ the vdW-Casimir free energy can be written as
a linear function of $N$ so that the interaction free energy
$F_N(a,b)$ becomes \cite{Rudi-JCP-124-044709-2006}
\begin{equation}
F_N(a,b)=N f_{gg}(a,b)\quad \qquad N\gg 1, \label{eq:Free-Energy-Many-Body-N-sheets}
\end{equation}
where $f_{gg}(a,b)$ can be interpreted as an {\em effective} pair interaction between
two neighboring layers in the stack and is given by
\begin{eqnarray}
f_{gg}(a,b) &=& k_B T \times
\nonumber\\
&& \hspace{-1.8cm}  \times \sum_{Q} {\sum_{n=0}^{\infty}}^\prime \ln \frac{1}{2} \bigg[
\frac{1- \Delta^2 ( e^{-2 \kappa_A a } + e^{-2 \kappa_B b }  )+e^{-2 (
\kappa_A a  + \kappa_B b ) }}{(1 - \Delta^2 e^{-2 \kappa_A a } )}
\nonumber\\
&& \hspace{-0.1cm} + \sqrt{\frac{{\cal{G}} (a,b,\Delta)}{(1-
\Delta^2 e^{-2 \kappa_A a } )^2}} \bigg],
\label{eq:Att-vdW-sheets-per-number-of-layers}
\end{eqnarray}
with ${\cal{G}} (a,b,\Delta)$ defined as
\begin{eqnarray}
{\cal{G}}(a,b,\Delta) &=&\big( 1-e^{-2 ( \kappa_A a + \kappa_B b ) }
\big)^2 \nonumber\\
&-& 2 \Delta^2 \bigg[( e^{-2 \kappa_A a } + e^{-2 \kappa_B b })(1+ e^{-2 (\kappa_A a + \kappa_B b )}) \nonumber\\
&-& 4 e^{-2 (\kappa_A a + \kappa_B b )} \bigg] + \Delta^4 \big( e^{-2
\kappa_A a} - e^{-2
\kappa_B b}  \big)^2. \nonumber\\
\label{eq:cal-G}
\end{eqnarray}
Here $\Delta$ is
\begin{equation}
\Delta = \frac{\kappa_A \epsilon_B - \kappa_B \epsilon_A}{\kappa_A
\epsilon_B + \kappa_B \epsilon_A },  
\end{equation}
where $\kappa_A$ and $\kappa_B$ are
\begin{equation}
\kappa_{A,B}^2= Q^2 +  \xi_n^2 ~\epsilon_{A,B} (\imath \xi_n)
/c^2 . \label{eq:rho(A,B)}
\end{equation}
For simplicity we have dropped the explicit dependence on the Matsubara frequency in all the above expressions.
We stress that $f_{gg}(a,b)$ is an effective pair interaction between two neighboring graphene layers in a multilayer geometry
and is in general {\em not} equal to $F_{gg}(a, b)$ in Eq. (\ref{eq:Free-Energy-Two-sheets}), which is valid for two interacting layers in the absence of any other neighboring layers. The difference between these two interaction free energies thus encodes the non-pairwise additive effects in the interaction between two layers due to the presence of other vicinal layers.

%
\begin{figure}[t]
\begin{center}
\includegraphics[width=8.5cm]{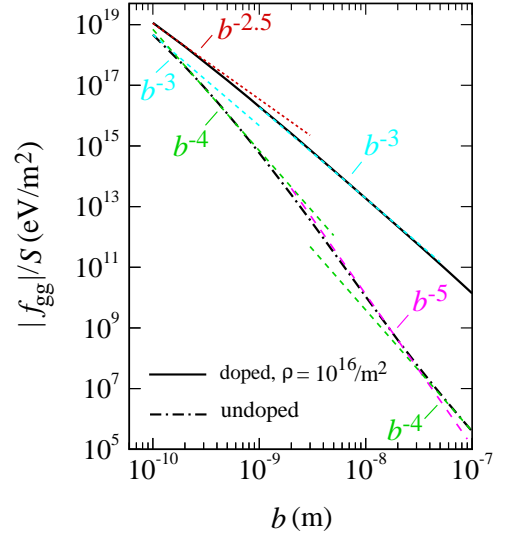}
\caption{ (Color online) Magnitude of
the interaction free energy per unit area and number of
layers, $|f_{gg}|/S$, plotted as a function of the separation, $b$, 
between two successive undoped (bottom black dot-dashed line) and doped
(top black solid line) graphene layers for a system composed of
infinitely many layers of undoped/doped layers as schematically depicted in
Fig.~\ref{fig:N-layer-New}. The temperature of the system is
$300~\mathrm{K}$, the thickness of the graphene layers is fixed
at $1$\AA~and we have chosen the dielectric function
for the undoped case from
Eq.~(\ref{eq:epsilon-undoped-graphene-sheet}) and for doped
case (with the doping electron density  chosen as $\rho=10^{16}
/\mathrm{m}^2$) from
Eq.~(\ref{eq:dielectric-function-Doped-graphene-sheet}). The
functional  form of the  free energy for the undoped case is compared with scaling forms $b^{-3}, b^{-4}, b^{-5}$
and $b^{-4}$ in various regimes of separation. For the doped case, 
the free energy is compared with the scaling forms 
 $b^{-2.5}$ and $b^{-3}$.} \label{fig:Energy-Doped-and-undoped-N_Layers-per-area-number-of-layer}
\end{center}
\end{figure}

\subsection{$N+1$ undoped graphene layers}

Let us first consider the case of $N+1$ undoped
graphene layers. In this case the vdW-Casimir interaction free energy per unit
area and per number of layers, $|f_{gg}(b)|/S$ (black dot-dashed line), has
been plotted in Fig.~\ref{fig:Energy-Doped-and-undoped-N_Layers-per-area-number-of-layer}
as a function of the separation between the  layers, $b$. The
temperature of the system is chosen as $300~\mathrm{K}$, the
thickness of the graphene layers is $1$\AA~and we
have used the dielectric function given by Eq.~(\ref{eq:epsilon-undoped-graphene-sheet}) for each undoped
graphene sheet.

The value of $|f_{gg}|/S$ at $1~\mathrm{nm}$ is about
$5.9 \times 10^{14}~\mathrm{eV}/\mathrm{m}^2$ which is
$5.9\times10^{2}~\mathrm{eV}$ when the surface area is equal to
$10^{-12}~\mathrm{m}^2$. At $10~\mathrm{nm}$ the value of
$|f_{gg}|$ for the same surface area is about $0.01~\mathrm{eV}$ and
at $100~\mathrm{nm}$ is about $3.9 \times 10^{-7}
~\mathrm{eV}$.

The scaling of $|f_{gg}(b)|/S$ for different values of the interlayer spacing $b$
is shown in Fig.~\ref{fig:Energy-Doped-and-undoped-N_Layers-per-area-number-of-layer}.
It shows the scaling exponent $n = 3$ at vanishing separations while at finite yet small separations
it is characterised by $n = 4$, continuously merging into a $n = 5$ form and finally attaining the $n = 4$ form.
The rationalization of this sequence of scaling exponents is exactly the same as in the case of two isolated
layers and will thus not be repeated here.

\begin{figure}[t]
\begin{center}
\includegraphics[width=8.5cm]{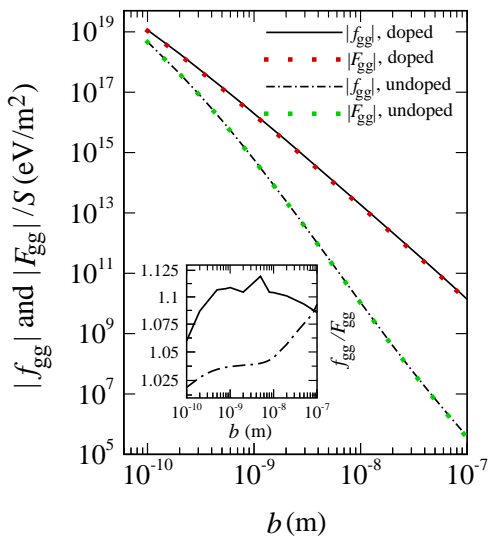}
\caption{ (Color online) Magnitude of the interaction free energy per unit area and number of
layers, $|f_{gg}|/S$, for the system composed of $N+1$ layers of undoped (bottom black dot-dashed
line) 
and doped (top black solid line) 
graphene is compared with the magnitude of the interaction free energy per unit
area, $|F_{gg}|/S$, for the system composed of only two undoped (green
dots) 
and doped
(red dots) 
graphene
layers as a function of the separation between the layers,  
$b$. The inset shows the ratio of
these two quantities for both undoped (black dot-dashed line) and doped (black solid
line) cases. }
\label{fig:Energy-NLayer-over-TwoLayer-Doped-and-Undoped}
\end{center}
\end{figure}

One can directly compare the reduced free energy, $f_{gg}(b)$, with that of a system composed of only two undoped
graphene layers of the same thickness $a$, $F_{gg}(b)$ (i.e., comparing the results in
Fig.~\ref{fig:Energy-two-undoped-graphene-sheet} with the corresponding results in 
Fig.~\ref{fig:Energy-Doped-and-undoped-N_Layers-per-area-number-of-layer}).
This is shown in Fig.~\ref{fig:Energy-NLayer-over-TwoLayer-Doped-and-Undoped}
(black dot-dashed line and green dots), where apparently
the results nearly coincide.  However, by inspecting the ratio
between these two interaction free energies it turns out that in the multilayer
system the interaction free energy per layer is slightly more
attractive than in the case of a two-layer system. This difference thus stems directly
from the many-body effects which in this case augment the binding interaction in
a graphitic stack. 


%

\subsection{$N+1$ doped graphene layers}

In Fig.~\ref{fig:Energy-Doped-and-undoped-N_Layers-per-area-number-of-layer}
the magnitude of the vdW-Casimir interaction free energy for doped
graphene layers per unit area and per number of layers,
$|f_{gg}(b)|/S$, is plotted (black solid line) as a function of the
separation between layers, $b$. The vdW-London dispersion
transform of the dielectric function for each graphene sheet is
chosen as in
Eq.~(\ref{eq:dielectric-function-Doped-graphene-sheet}). We have
fixed the density of electrons for all the graphene layers as
$\rho = 10^{16} /\mathrm{m}^2$. The value of $|f_{gg}|/S$ at
$1~\mathrm{nm}$ is about $1.96\times 10^{16}~\mathrm{eV}/\mathrm{m}^2$, which is about $1.96\times10^{4}~
\mathrm{eV}$ when the surface area is equal to
$10^{-12}~\mathrm{m}^2$. At $10~\mathrm{nm}$ the value of
$|f_{gg}|$ is about $18.7~\mathrm{eV}$ and at $100~\mathrm{nm}$ it
is about $1.38 \times 10^{-2} ~\mathrm{eV}$. The scaling
exponents of the interaction free energy dependence for different
regions of interlayer spacings are illustrated in
Fig.~\ref{fig:Energy-Doped-and-undoped-N_Layers-per-area-number-of-layer}.
The scaling exponent is  $n = 2.5$ for small separations while at
larger separations it tends towards the value $n = 3$. It is thus
exactly the same as in the case of two isolated doped graphene
sheets, see Fig.~\ref{fig:Energy-two-Doped-graphene-sheet},
except that in the multilayer geometry we have not shown the same
range of separations as for two isolated layers.

The comparison of  vdW-Casimir interaction free energies in the case of two
isolated doped graphene sheets with the effective interaction between two graphene sheets in a multilayer system
(i.e., comparing $|F_{gg}|$ in Fig.~\ref{fig:Energy-two-Doped-graphene-sheet} with $|f_{gg}|$ in
Fig.~\ref{fig:Energy-Doped-and-undoped-N_Layers-per-area-number-of-layer})
is made  in Fig.~\ref{fig:Energy-NLayer-over-TwoLayer-Doped-and-Undoped}. As
shown by the inset the interaction free energy is again slightly more attractive within a
multilayer. Comparing the results in the inset of Fig.~\ref{fig:Energy-NLayer-over-TwoLayer-Doped-and-Undoped}
shows that in average the many-body effects are stronger in the
doped multilayer case than in the undoped case.

Note also that a direct comparison between the undoped and
doped systems (Fig.~\ref{fig:Energy-Doped-and-undoped-N_Layers-per-area-number-of-layer})
shows that for all separations the free energy magnitude for a doped multilayer (solid line) is
more than that of the undoped one (dot-dashed line) and thus the interaction is more attractive in the former case. 
At separation $1~\mathrm{nm}$ the magnitude of the doped free energy is about
34 times larger, while at the separation $10~\mathrm{nm}$ it is
about 1730 larger than that of the undoped one.

\section{Conclusion}

In this work we have studied the vdW-Casimir interaction between graphene sheets and between a
graphene sheet and a substrate. We calculated the interaction free energy via the Lifshitz theory of vdW
interactions that takes as an input the dielectric functions, or better their vdW-London transform, of isolated
layers. Within this approach it would be inconsistent to take into account any separation dependent coupling between
the dielectric  response of the layers. This need possibly not be the case for some other approximate approaches to vdW
interactions as in, e.g., the vdW augmented density functional theory (see the paper by Langreth et al. in Ref.~\cite{vdWgeneral}).

By inserting the {\em random phase approximation} dielectric function of a graphene layer into the Lifshitz theory we are thus in a position to evaluate not only the pair interaction between two isolated graphene sheets, but also between a graphene sheet and a
semi-infinite substrate of a different dielectric nature (SiO$_2$ in our case) as well as the effective interactions between
two graphene sheets in an infinite stack of graphene layers. All these cases that have been analyzed and discussed above are relevant for many realistic geometries in nano-scale systems~\cite{Rudi-RMP-2010} and thus deserve to be studied in detail.

In the three cases studied we found the following salient features of the vdW-Casimir interaction dependence on the separation between the interacting bodies:
\begin{itemize}
\item[1-] In a system composed of two graphene layers we demonstrated that the vdW-Casimir interactions in the case of undoped graphene show scaling exponents identical to those displayed in the case of interacting thin dielectric layers. In the doped case the scaling exponents are consistent with vdW-Casimir interactions between two thin metallic layers.
\item[2-] In a system composed of a semi-infinite dielectric substrate and an undoped graphene layer the vdW-Casimir interactions display scaling exponents expected for this asymmetric geometry. For a doped graphene layer the exponents revert to the previous case of two doped graphene layers.
\item[3-] In a multilayer system composed of many graphene sheets the vdW-Casimir interaction scaling exponents are the same as in the case of two isolated layers but the interactions are stronger due to many body effects as a consequence of the presence of other layers in a stack.
\end{itemize}

In order to describe the correlation effects especially at low doping or the interlayer coupling 
on a more systematic level, one needs to go beyond the standard random phase approximation
by incorporating more sophisticated theoretical models for the dielectric response function which would be worth 
exploring further in the future.

The main motivation for a detailed study of vdW-Casimir interaction between graphene sheets in graphite-like geometries is the fact that graphitic systems belong to closed shell systems and thus display  no covalent bonding, so that any bonding interaction is by necessity of a vdW-Casimir type. Its detailed characterization is thus particularly relevant for this quintessential nano-scale system \cite{Rudi-RMP-2010}.

\section{Acknowledgment}

We would like to thank B. Sernelius for providing us with a
preprint of his work. R.P. acknowledges support from ARRS through
the program P1-0055 and the research project J1-0908. A.N.
acknowledges support from the Royal Society, the Royal Academy of
Engineering, and the British Academy. J.S. acknowledges generous
support by J. Stefan Institute (Ljubljana) provided for a visit to
the Institute and the Department of Physics of IASBS (Zanjan) for their 
hospitality.


\end{document}